\documentclass[twocolumn,aps,showpacs]{revtex4}
\usepackage{epsfig}

\makeatother
\newcommand{\boldsymbol}[1]{\mbox{\boldmath $#1$}}

\begin{document}

\title{Linear spin waves in a trapped Bose gas}

\author{T. Nikuni}
\altaffiliation[Present address: ]{Department of Physics,
Faculty of Science, Tokyo University of Science,
1-3 Kagurazaka, Shinjuku-ku, Tokyo 162-8601, Japan.}
\affiliation{Department of Physics, University of Toronto,
Toronto, Ontario, Canada M5S 1A7}
\author{J. E. Williams}
\author{C. W. Clark}
\affiliation{Electron and Optical Physics Division, National
Institute of Standards and Technology, Gaithersburg, Maryland
20899-8410}

\date{\today}

\begin{abstract}
\vspace{.2cm} An ultra-cold Bose gas of two-level atoms can be thought
of as a spin-1/2 Bose gas.  It supports spin-wave collective modes due
to the exchange mean field.  Such collective spin oscillations have
been observed in recent experiments at JILA with ${}^{87}$Rb atoms
confined in a harmonic trap. We present a theory of the spin-wave
collective modes based on the moment method for trapped gases. In the
collisionless and hydrodynamic limits, we derive analytic expressions
for the frequencies and damping rates of modes with dipole and
quadrupole symmetry. We find that the frequency for a given mode is
given by a temperature independent function of the peak density $n$,
and falls off as $1/n$.  We also find that, to a very good
approximation, excitations in the radial and axial directions are
decoupled. We compare our model to the numerical integration of a one
dimensional version of the kinetic equation and find very good
qualitative agreement.  The damping rates, however, show the largest
deviation for intermediate densities, where one expects Landau damping
-- which is unaccounted for in our moment approach -- to play a
significant role.
\end{abstract}


\maketitle

%\bigskip

\section{Introduction}
Collective spin oscillations are a general consequence of the quantum
exchange between identical particles in a system where a macroscopic
symmetry breaking exists in spin space~\cite{Ashcroft}. From a
condensed matter perspective, spin waves are most familiar in strongly
interacting degenerate Fermi systems, such as a ferromagnet. It is
somewhat counter-intuitive, although well established both
experimentally~\cite{Johnson84,Bigelow89,Gully84} and theoretically
\cite{Leggett68,Bashkin81,Lhuillier1,Levy84,Ruckenstein89}, that
collective spin behavior can also occur in nondegenerate dilute spin
polarized gases when the thermal de Broglie wavelength exceeds the
effective range of interaction between two colliding atoms. In these
systems, a transverse spin wave is excited by applying an
inhomogeneous magnetic field followed by a small tipping pulse. It is
remarkable that the mean field generates collective spin dynamics, but
has no discernible effect on thermodynamic equilibrium properties
since $gn/k_BT\ll 1$, where $g$ is the binary interaction parameter,
$n$ is the density and $T$ the temperature.

Recent experiments at JILA~\cite{Lewandowski,McGuirk} on a trapped
$^{87}$Rb gas have revived interest in spin waves in dilute
gases~\cite{Oktel,Fuchs,Williams,Kuklov}. These new experiments offer
several interesting new features compared to the earlier experiments
in spin-polarized hydrogen~\cite{Johnson84,Bigelow89}, which take
advantage of the technological advances made over the last twenty
years in the measurement and control of cold atomic gases. A prominent
feature of the new generation of experiments is the ability to take
spatially resolved measurements of the gas sample using absorption
imaging techniques. Another exciting advancement is the ability to
cool the sample into the quantum degenerate regime, which permits the
study of spin waves in a Bose-condensed gas at finite temperatures;
this regime has never been investigated experimentally and has only
received minor attention in the theoretical
literature~\cite{Oktel2,Oktel3,NW}. In this paper, however, we focus
on the noncondensed regime relevant to the recent JILA experiments
\cite{Lewandowski,McGuirk}, where the temperatures are approximately
twice that needed for Bose-Einstein condensation $T_{\rm{BEC}}$.

The JILA system consists of a dilute gas of ${}^{87}$Rb atoms that
have been optically pumped into the
$|F=1,M_F=-1\rangle\equiv|1\rangle$ hyperfine state and are confined
in a magnetic harmonic trapping potential. By applying microwave and
radio-frequency radiation that couples to the
$|2,1\rangle\equiv|2\rangle$ state, atoms in the gas can be uniformly
prepared in an arbitrary superposition of the $|1\rangle$ and
$|2\rangle$ states. This system can be thought of as a spin-1/2 system
by taking $|1\rangle$ as the spin-up state and $|2\rangle$ as the
spin-down state. In Figure 1 we illustrate the corresponding Bloch
vector spin describing the internal state of the atoms. Note that
because the magnetic field direction varies in the
trap~\cite{Bergeman}, the spin axis shown in Figure 1 is not
isomorphic with the coordinate axis describing the position of an
atom, in contrast to the situation in spin-polarized hydrogen.
\begin{figure}
  \centerline{\epsfig{file=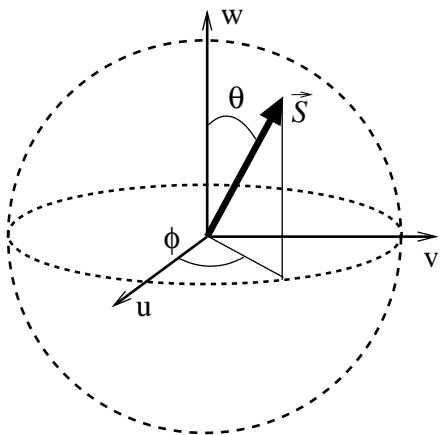,height=2.25in}}
\caption{Schematic diagram of the Bloch vector spin $\vec{S}$. The
spins of the atoms are initially pointing up, $\vec S = \{0,0,S_w\}$,
corresponding to all the atoms being in the state $|1\rangle$. After
the $\pi/2$ pulse, the spin vector is pointing along the $v$ axis,
$\vec{S}=\{0,S_v,0\}$, corresponding to the state $(|1\rangle + i
|2\rangle)/\sqrt{2}$.}
\end{figure}

In the JILA experiment, an initial $\pi/2$ pulse is applied to tip the
spins into the transverse direction $v$. The spin vector then
precesses about the longitudinal $w$-axis at a rate proportional to
the energy difference between hyperfine states. Due to the mean field
and differential Zeeman effects, the local frequency splitting between
hyperfine states varies approximately quadratically with position.
This inhomogeneity initiates collective spin dynamics through the
exchange mean field, the initial onset of which gives the striking
appearance of spin
segregation~\cite{Lewandowski,Oktel,Fuchs,Williams}. The inhomogeneous
frequency splitting can be made arbitrarily small to study the linear
response of the system.  Recently this technique was used to probe
intrinsic collective spin oscillations~\cite{McGuirk}.  Studies of
such spin-wave collective modes provide us clearer physical
understanding of the important role of the exchange interactions in a
dilute Bose gas.

The experiments described in Refs.~\cite{Lewandowski,McGuirk} present
spatially resolved images of spin dynamics in a gas. The density
profile of either state is measured using absorption imaging.
Together with the Ramsey fringe technique, integrated spatial profiles
of the longitudinal spin $S_w$ and transverse phase $\phi$ can be
extracted from experimental data, as shown in the stunning images in
Ref.~\cite{McGuirk}. This is in sharp contrast to the earlier hydrogen
experiments, where a pulsed NMR technique is used to obtain the
frequency of spin oscillation integrated over the entire sample. In
this paper we show a theoretical prediction of the spatial structure
of the spin dynamics that qualitatively agrees with experimental
observation.

In this paper, we present the theory of spin waves in a trapped Bose
gas using the moment method \cite{GO,Khawaja}, which was originally
developed to study collective density oscillations in a trapped
classical gas.  The moment approach has also been applied to a
rotating gas \cite{GO2} and was recently generalized to treat a
Bose-condensed gas at finite temperature to study the scissors modes
\cite{Nikuni}.  An advantage of this technique is that the solution
maps smoothly between the collisionless and hydrodynamic regimes
\cite{GO}.  We apply the moment method to a spin kinetic equation, and
derive explicit analytical expressions for frequency and damping of
dipole and quadrupole modes in weak and strong coupling limits.  We
also numerically solve a one-dimensional model of the kinetic
equation, and compare with the moment results.

\section{Spin kinetic equations}
The Hamiltonian describing a single, trapped, two-level atom of mass
$m$ is:\begin{equation}
\label{H1}
\hat{H}=\left[-\frac{\hbar ^{2}}{2m}\nabla
^{2}+U_{\mathrm{ext}}(\mathbf{r}) \right]\hat{1}+\frac{\hbar
}{2}\vec{\Omega }(\mathbf{r})\cdot \vec{\hat{\tau }}.
\end{equation}
The first term in Eq.~(\ref{H1}) is the center of mass Hamiltonian
containing the kinetic energy and the external parabolic trap
potential
\begin{equation}
U_{\mathrm{ext}}(\mathbf{r})= {m\omega _{z}^{2}\over 2} \left[ \alpha
^{2}(x^{2}+y^{2})+z^{2}\right] , 
\end{equation}
where \( \alpha =\omega _{r}/\omega _{z} \). This part of the
Hamiltonian is uncoupled from the internal, pseudo-spin, degree of
freedom, which is governed by the second term: \( \vec{\Omega
}(\mathbf{r})\cdot \vec{\hat{\tau }}=\Omega _{u}(\mathbf{r})\hat{\tau
}_{u}+\Omega _{v}(\mathbf{r})\hat{\tau }_{v}+\Omega
_{w}(\mathbf{r})\hat{\tau }_{w}, \) where \( \hat{\tau }_{i} \) is a
Pauli matrix.  In the absence of an external coupling field,
$\Omega_u=\Omega_v=0$ and
$\Omega_w=\Delta(\mathbf{r})$~\cite{Lewandowski,McGuirk} is the
frequency splitting between the two states (we go to a rotating frame
to eliminate the large hyperfine splitting $\omega_{\mathrm{hf}}$
frequency). We model binary interactions between particles by a
delta-function pseudo potential describing elastic, spin preserving
collisions, the strength of which depends on the hyperfine states
$V_{ij}(\mathbf{r},\mathbf{r}')= g_{ij}\delta
(\mathbf{r}-\mathbf{r}')$, where $g_{ij}=4\pi \hbar ^{2}a_{ij}/m$,
with $a_{ij}$ being the scattering length for collisions between atoms
of species $i$ and $j$.  For ${}^{87}$Rb, \( a_{11}=100.9a_{0} \), \(
a_{12}=98.2a_{0} \), \( a_{22}=95.6a_{0} \), where \( a_{0} \) is the
Bohr radius~\cite{Lewandowski}. In the rest of this paper, however, we
make the simplification that $a_{11}=a_{22}=a_{12}\equiv a$, which is
a reasonable approximation for ${}^{87}$Rb.

Several groups have previously worked out the fundamental kinetic
theory of a noncondensed dilute Bose gas with internal degrees of
freedom, to describe spin waves in spin-polarized atomic hydrogen
above $T_{\rm{BEC}}$~\cite{Bashkin81,Lhuillier1,Levy84,Ruckenstein89}.
Using a semiclassical approximation to describe atomic motion in terms
of a phase-space distribution function, we obtain coupled Boltzmann
equations for the atomic
\( f(\mathbf{r},\mathbf{p},t) \) and spin \( \vec{\sigma
}(\mathbf{r},\mathbf{p},t) \) distribution functions:
\begin{equation}
\label{kineticfcm}
\frac{\partial f}{\partial t}+\frac{\mathbf{p}}{m}\cdot 
\boldsymbol \nabla_r f-\boldsymbol \nabla U_{n}\cdot 
\boldsymbol \nabla _{p}f
-\frac{\hbar }{2}\boldsymbol \nabla \Omega _{ni }\cdot
\boldsymbol \nabla _{p}\sigma _i
=\left. \frac{\partial f}{\partial t}\right| _{\mathrm{coll}} ,
\end{equation}
\begin{equation}
\label{kineticfs}
\frac{\partial \vec{\sigma }}{\partial t}+\frac{\mathbf{p}}{m}\cdot
\boldsymbol \nabla_r \vec{\sigma }-\boldsymbol \nabla U_{n}\cdot
\boldsymbol \nabla _{p}\vec{\sigma }-\frac{\hbar }{2}\boldsymbol
\nabla \vec{\Omega }_{n}\cdot \boldsymbol \nabla _{p}f-\vec{\Omega
}_{n}\times \vec{\sigma }=\left. \frac{\partial \vec{\sigma
}}{\partial t}\right| _{\mathrm{coll}} .
\end{equation}
Equation~(\ref{kineticfcm}) has an implicit sum over the repeated
index $i=u,v,w$ in the fourth term.  The total density and spin
density are obtained from the distribution functions as \(
n(\mathbf{r},t)\equiv n_1({\bf r},t)+n_2({\bf r},t)= \int
d\mathbf{p}f(\mathbf{r},\mathbf{p},t)/(2\pi \hbar )^{3} \) and \(
\vec{S}(\mathbf{r},t)= \int
d\mathbf{p}\vec{\sigma}(\mathbf{r},\mathbf{p},t)/(2\pi \hbar)^{3} \)
respectively. Here the longitudinal component of the spin represents
the relative density $S_w=n_1-n_2$ and the transverse components $S_u$
and $S_v$ describe the real and imaginary parts of the internal
coherence.  The center of mass effective potential is \(
U_{n}(\mathbf{r},t)=U_{\mathrm{ext}}(\mathbf{r})+ 3 g n({\bf r},t) / 2
.\) The effective coupling field including mean-field effects is
\begin{equation}
\label{Omegan}
\vec{\Omega }_{n}(\mathbf{r},t)=
\vec{\Omega }(\mathbf{r},t)+{g\over \hbar }\vec{S}(\mathbf{r},t).
\end{equation}
The collision integrals in Eq.~(\ref{kineticfcm}) and
Eq.~(\ref{kineticfs}) are written explicitly in the Appendix.

The center of mass and spin are coupled via the third and fourth terms
in both Eq.~(\ref{kineticfcm}) and Eq.~(\ref{kineticfs}). These terms
involve spatial gradients over the density and scale like $g n / k_B
T\ll 1$, and thus can be neglected. This allows us to make the further
simplification that the center of mass and spin dynamics are
decoupled. Since we are interested in the intrinsic collective modes
of the system, we also assume that the field
$\vec\Omega(\mathbf{r})=(0,0,\Delta(\mathbf{r}))$ can be made to
vanish after the spin wave is excited. This assumption is motivated by
the JILA experiment~\cite{McGuirk}, where the static inhomogeneous
frequency splitting could be adjusted to zero after a short excitation
time. The position dependence of $\Delta(\mathbf{r})$ determines the
symmetry of collective mode excited. With these two assumptions, the
kinetic equation for the spin distribution function then simplifies to
\begin{equation}
\label{kineticfs2}
\frac{\partial \vec{\sigma }}{\partial t}+\frac{\mathbf{p}}{m}\cdot
\boldsymbol \nabla \vec{\sigma }-\boldsymbol \nabla U_{\rm{ext}}\cdot
\boldsymbol \nabla _{p}\vec{\sigma
}-\frac{g}{\hbar}\vec{S}\times\vec{\sigma}=\left. \frac{\partial
\vec{\sigma }}{\partial t}\right| _{\mathrm{coll}} .
\end{equation}

In this paper, we consider small amplitude spin oscillations around a
fully-polarized state. It is convenient to define new spin coordinates
$(u',v',w')$ with $w'$ being the direction of the initial spin
polarization.  In the experiment described in Ref.~\cite{McGuirk}, the
spin oscillations occur about the spin state polarized along the $v$
axis. In this case one must make the mapping of the spin coordinates
$(u',v',w')=(w,u,v)$, corresponding to a cyclic
permutation~\footnote{In this paper, the directions ``transverse'' and
``longitudinal'' are defined relative to the direction of polarization
of the initial spin distribution. Therefore, these terms have a
different meaning than in~\cite{Williams}, where the terms are defined
relative to the quantization axis of the spin.}. We then linearize the
kinetic equation around the equilibrium state polarized along the $w'$
direction (we drop the prime from our notation from here on):
$\sigma_{u0}=\sigma_{v0}=0$, and $\sigma_{w0}({\bf r},{\bf
p})=f_0({\bf r},{\bf p})$, where the equilibrium distribution is
\begin{equation}
f_0({\bf r},{\bf p}) =\exp\left\{-\beta\left[\frac{p^2}{2m}+U_{\rm
ext}({\bf r})-\mu_0\right]\right\}.
\label{f0}
\end{equation}
The equilibrium density given from Eq.~(\ref{f0}) is 
\begin{equation}
n_0({\bf{r}})=\int \frac{d{\bf{p}}}{(2\pi\hbar)^3}f_0({\bf r},{\bf p})
=\frac{1}{\lambda_{\rm th}^3}\exp[-\beta U_{\rm ext}({\bf r})],
\label{n0}
\end{equation} 
where $\lambda_{\rm th}=(2\pi\hbar^2/mk_{\rm B}T)^{1/2}$ is the
thermal de Broglie wavelength.
We then substitute $\vec{\sigma }(\mathbf{r},\mathbf{p},t) =
\vec{\sigma }_0(\mathbf{r},\mathbf{p}) + \delta \vec{\sigma
}(\mathbf{r},\mathbf{p},t)$ into (\ref{kineticfs2}) to obtain the
linearized spin kinetic equation
\begin{eqnarray}
\frac{\partial\delta\vec{\sigma}}{\partial t}&+& \frac{{\bf
p}}{m}\cdot\boldsymbol \nabla\delta\vec{\sigma} -\boldsymbol \nabla
U_{\rm ext}\cdot\boldsymbol \nabla_p\delta\vec{\sigma} \nonumber \\
&-&\frac{g}{\hbar}(\vec{S}_0\times\delta\vec{\sigma} +\delta
\vec{S}\times\vec{\sigma}_0)=\left.\frac{\partial\vec{\sigma}}{\partial
t} \right|_{\rm coll}.
\label{linear_eq}
\end{eqnarray}
The linearized form of the collision integral is discussed in the
Appendix.

The longitudinal spin dynamics is described by
\begin{equation}
\frac{\partial\delta\sigma_w}{\partial t}+\frac{{\bf
p}}{m}\cdot\boldsymbol \nabla \delta\sigma_w-\boldsymbol \nabla U_{\rm
ext}\cdot\boldsymbol \nabla _p\delta\sigma_w
=\left.\frac{\partial \delta \sigma_w}{\partial t}\right|_{\rm coll}.
\label{longitudinal}
\end{equation}
Since the mean field term does not appear in (\ref{longitudinal}),
collective oscillations of the longitudinal spin only occur in the
low-density collisionless regime in a trapped gas.  In the
high-density hydrodynamic regime, longitudinal modes become purely
relaxational modes damped by the diffusion transport process.  The
crossover from the collisionless spin oscillation to the hydrodynamic
relaxation mode in a longitudinal spin excitation was observed in a
trapped $^{40}$K fermi gas \cite{Jin1,Jin2} (although they were not
excitations from the fully polarized state as considered here).  In
contrast, transverse spin waves behave collectively due to the mean
field and will be the focus of this paper.  For the transverse spin
fluctuations, it is convenient to work with
$\delta\sigma_{\pm}\equiv\delta\sigma_u\pm i\delta\sigma_v$.  We then
obtain
\begin{eqnarray}
\frac{\partial\delta\sigma_{\pm}}{\partial t}&+& \frac{{\bf
p}}{m}\cdot\boldsymbol \nabla \delta\sigma_{\pm}- \boldsymbol \nabla
U_{\rm ext}\cdot\boldsymbol \nabla _p\delta\sigma_{\pm} \nonumber \\
&\pm& i\frac{g}{\hbar}(f_0\delta S_{\pm}-n_0\delta\sigma_{\pm})
=\left.\frac{\partial\delta\sigma_{\pm}} {\partial t}\right|_{\rm
coll}.
\label{eq_sigmapm}
\end{eqnarray}

Although we are interested in a trapped Bose gas,
it is useful to summarize earlier results on the theory of spin waves
in a homogeneous gas \cite{Smith}. 
In the long wavelength limit, where
the gas can be treated in the hydrodynamic regime, the dispersion
relation has the form~\cite{Levy84,Smith}
\begin{equation}
\omega(k) = -i k^2 v_{\rm{th}}^2 \tilde\tau_D
\label{omegak1}
\end{equation}
where $v_{\rm{th}}=\sqrt{k_B T/m}$ and $\tilde\tau_D$ is a complex
diffusive relaxation time. For transverse spin oscillations, one finds
$\tilde\tau_D^{-1}=(\tau_D^{-1} - i g n/\hbar)$, where the diffusive
relaxation time is $\tau_D = [(32a^2 n /3)\sqrt{\pi k_B
T/m}]^{-1}$. Due to the exchange mean field, the transverse spin
behaves collectively, and in the limit where $gn \tau_D /\hbar \gg 1$,
the dispersion relation has the form
\begin{equation}
\omega(k) = {\hbar^2 k^2 \over{2 m^*} } - {i \over \tau_D} \Big (
{ \hbar k v_{\rm{th}} \over {g n}} \Big )^2 ,
\label{omegak2}
\end{equation}
where $m^* = (gn/2 k_B T)m$ is regarded as an effective mass.
The $k^2$ dispersion relation is a universal result for ferromagnetic-like spin
systems~\cite{Ashcroft}.  The longitudinal spin oscillations do not
behave collectively, but rather exhibit a purely diffusive mode, with
$\tilde\tau_D\rightarrow\tau_D$ in Eq.~(\ref{omegak1}).

\section{Moment method for a trapped spin-$1/2$ gas}
We now turn to spin waves in a Bose gas confined in the harmonic trap potential.
Starting from Eq.~(\ref{eq_sigmapm}), one can derive a general set of coupled moment
equations associated with a set of polynomial functions $\chi_i({\bf
r},{\bf p})$:
\begin{eqnarray}
\frac{d\langle\chi_i\rangle}{dt} &-&\left\langle \boldsymbol \nabla
\chi_i\cdot\frac{{\bf p}}{m}\right\rangle +\langle\boldsymbol \nabla
U_{\rm ext}\cdot\boldsymbol \nabla _p\chi_i\rangle \nonumber \\
&+&i\frac{g}{\hbar}[\langle S_+\chi_i\rangle_0-\langle
n_0\chi_i\rangle] =\langle\chi_i\rangle_{\rm coll},
\label{mastermoment}
\end{eqnarray}
where the moment variables are defined as
\begin{equation}
\langle\chi_i\rangle\equiv\frac{1}{N}\int d{\bf r}\int
\frac{d{\bf p}}{(2\pi\hbar)^3} \chi_i({\bf r},{\bf
p})\delta \sigma_+({\bf r},{\bf p},t),
\label{momentdef}
\end{equation}
\begin{equation}
\langle\chi_i\rangle_{\rm coll}\equiv\frac{1}{N}\int d{\bf r}
\int \frac{d{\bf p}}{(2\pi\hbar)^3}\chi_i({\bf r},{\bf p})
 \left.\frac{\partial\delta \sigma_+}{\partial t}
\right|_{\rm coll},
\end{equation}
\begin{equation}
\langle\chi_i\rangle_0\equiv \frac{1}{N}\int d{\bf r}\int
\frac{d{\bf p}}{(2\pi\hbar)^3} \chi_i({\bf r},{\bf p}) f_0({\bf
r},{\bf p}).
\end{equation}
In general, moment equations are not closed since the mean field and
collisional terms couple to higher moments.  Those higher moments can
be truncated by expanding the fluctuations in the distribution
function $\delta\sigma_+$ in powers of position and momentum. One can
relate the coefficients in the expansion to the moments of the
distribution function $\langle \chi_i \rangle $ to yield a closed
set of equations that can be solved analytically.

The choice of the functions $\chi_i$ depends on the symmetry of the
spin-wave mode we are interested in.  In the following subsections, we
consider the dipole and quadrupole oscillations.

\subsection{Dipole mode}
We first consider the spin-wave collective mode with a dipole
symmetry, which could be excited by a linear inhomogeneous frequency
splitting, such as $\Delta(\mathbf{r})\propto z$. For our set of
moments to describe this oscillation, we choose:
\begin{equation}
\chi_1=x_i, ~\chi_2=p_{x_i}/m. 
\label{dipolemoments}
\end{equation}
The subscript indicates the axis along which the excitation oscillates
$x_i \in \{x,y,z \}$. Within the moment method approximate treatment,
the dipole modes along the three axis are completely decoupled.  We
also define $\chi_0=1$, related to the norm of $\delta \sigma_+$; we
set $\langle \chi_0 \rangle=0$, which is required from the
conservation of the spin density.

From Eq.~(\ref{mastermoment}), the equations of motion for the moments
Eq.~(\ref{dipolemoments}) are given by
\begin{eqnarray}
&&\frac{d}{dt}\langle \chi_1 \rangle=\langle\chi_2\rangle.
\label{eq_chi1_1} \\
&&\frac{d}{dt}\langle \chi_2\rangle+\omega_i^2\langle\chi_1\rangle
-i\frac{g}{\hbar}\langle n_0\chi_2\rangle=\langle\chi_2\rangle_{\rm coll},
\label{eq_chi2_1} 
\end{eqnarray}
where $\omega_i$ is either $\omega_z$ for an axial mode or
$\omega_\perp$ for a radial mode.  These are not a closed set of
equations, since in general the mean field and collisional terms in
Eq.~(\ref{eq_chi2_1}) couple to higher moments. The hierarchy of moment
equations can be truncated by assuming the explicit truncated form for
the distribution $\delta \sigma_+$:
\begin{equation}
\delta \sigma_+=f_0[\alpha_0 + \alpha_1 x_i + \alpha_2 p_{x_i}],
\label{dipoletruncate}
\end{equation}
The coefficients in the expansion can be related back to the set of
moments using (\ref{momentdef}): $\alpha_0=0$, $\alpha_1 = (m
\omega_i^2/k_B T) \langle \chi_1 \rangle$, and $\alpha_2 = \langle
\chi_2 \rangle /k_B T$.

Using the explicit form for $\delta \sigma_+$ in
Eq.~(\ref{dipoletruncate}), the mean-field and collisional terms can be
expressed in terms of the dipole moments. The resulting closed set of
coupled moment equations, written in matrix form, is
\begin{equation}
{d \over {dt}} \chi = \hat W_d \chi,
\label{dtWd}
\end{equation}
where the coupling matrix $\hat W_d$ is 
\begin{equation}
\label{Wd}
\hat W_d=-\left( \begin{array}{cc}
0 & -1\\
\omega _{i}^{2} & (\gamma _{D}-i\omega _{\mathrm{MF}})
\end{array}\right).
\end{equation}
Here we have defined the vector of moments ${\bf{\chi}}\equiv \{
\langle \chi_1 \rangle, \langle \chi_2 \rangle\}$.  Three different
frequencies appear in $\hat W_d$: the trap frequency $\omega_i$ along
the axis of oscillation, the mean field frequency $\omega_{\rm{MF}}$
defined by the spatial average $\omega_{\rm MF}\equiv \int d{\bf
r}gn_0^2({\bf r})/N\hbar$
\begin{equation}
\omega_{\rm MF}=\frac{gn_0(0)}{2\sqrt{2}\hbar},
\end{equation}
and the spatially averaged diffusion relaxation rate $\gamma_D$, the
form of which is given in the discussion in the Appendix.

We now look for normal mode solutions $\chi=\chi_0 e^{-i\omega t}$. 
Substituting this into Eq.~(\ref{dtWd}) yields an eigenvalue equation
with two solutions. It is straightforward to show that the mode
frequencies $\omega$ obey the dispersion relation
\begin{equation}
\omega^2+i\tilde\gamma_D\omega-\omega_i^2=0,
\label{dipoledispersion}
\end{equation}
where $\tilde\gamma_D\equiv\gamma_D-i\omega_{\rm MF}$ is the effective
(complex) diffusion relaxation rate including the mean-field effect.
The solution is given by
\begin{equation}
\omega=\frac{1}{2}\left[-i\tilde\gamma_D\pm\sqrt{4\omega_i^2-
\tilde\gamma_D^2} \right].
\end{equation}
We shall consider two limiting cases to obtain the scaling behavior of
the frequency and damping of the modes. In the weak interaction, or
collisionless, limit where $\omega_i\gg|\tilde\gamma_D|$, one has
\begin{equation}
\omega\simeq \pm\omega_z-\frac{\omega_{\rm MF}}{2}
-i\frac{\gamma_D}{2}.
\label{dipoleweak}
\end{equation}
In the strong coupling, or hydrodynamic, limit where
$\omega_i\ll|\tilde\gamma_D|$, one has
\begin{equation}
\omega \simeq \left\{ \begin{array}{c}
-i\omega _{i}^{2}/\tilde{\gamma }_{D}\, \, \, (\mathrm{low})\\
-i\tilde{\gamma }_{D}\, \, \, \, \, \, \, \, \, \, \, \, (\mathrm{high})
\end{array}\right.
\end{equation}
These represent low and high frequency modes, with the low frequency
mode having a higher $Q\sim {\rm {Re}}\omega / {\rm{Im}}\omega$
value. The low-frequency solution has the form of the diffusion
relaxation rate with a complex diffusion coefficient.  More
explicitly, the dispersion relation in the strong-coupling limit takes
the approximate form
\begin{equation}
\omega\simeq{\omega_i^2 \over {\omega_{\rm{MF}}}} \Big ( 1 - 
{\gamma_D^2 \over \omega_{\rm{MF}}^2}-i
{\gamma_D \over \omega_{\rm{MF}}} \Big ) .
\label{dipolestrong}
\end{equation}
The scaling behavior of the real part of Eq.~(\ref{dipolestrong}) can
be recovered in a simple model based on the homogeneous gas result of
Eq.~(\ref{omegak1}). For low-frequency collective modes in a trapped
gas along the $x_i$ direction, the wave vector $k$ is estimated as
$k\sim 1/R_i$, where $R_i=m\omega_i/k_{\rm B}T$ is the size of the
cloud along the $x_i$ direction.

\subsection{Quadrupole mode}
In the JILA experiment~\cite{McGuirk}, a spin-wave collective mode
with a quadrupole symmetry is excited due to an approximately
quadratically varying frequency splitting $\Delta(\mathbf{r})\propto
z^2$. In principle, the oscillation may be excited along both the
axial or radial directions, and so we take the following quantities
for our set of moments:
\begin{eqnarray}
&&\chi_1=z^2,~\chi_2=zp_z/m,~\chi_3=p_z^2/m^2, \\
&&\chi_4=r_{\perp}^2,~\chi_5={\bf r}_{\perp}\cdot{\bf p}_{\perp}/m,
~\chi_6=p_{\perp}^2/m^2,
\end{eqnarray}
where $r_\perp=\sqrt{x^2+y^2}$ and $p_\perp=\sqrt{p_x^2 + p_y^2}$. 
We also define $\chi_0=1$, related to the norm of $\delta \sigma$,
which we set to zero $\langle \chi_0 \rangle=0$.

The six moment equations for the above quantities are
\begin{eqnarray}
&& \frac{d\langle\chi_1\rangle}{dt}-2\langle\chi_2\rangle=0, \\
&& \frac{d\langle\chi_2\rangle}{dt}-\langle\chi_3\rangle
+\omega_z^2\langle\chi_1\rangle-i\frac{g}{\hbar}\langle n_0\chi_2\rangle
=\langle\chi_2\rangle_{\rm coll}, \\
&&\frac{d\langle\chi_3\rangle}{dt}
+2\omega_z^2\langle\chi_2\rangle+i\frac{g}{\hbar}
\left[\langle S_+\chi_3\rangle_0-\langle n_0\chi_3\rangle\right]
=\langle\chi_3\rangle_{\rm coll},\nonumber \\ \\
&& \frac{d\langle\chi_4\rangle}{dt}-2\langle\chi_5\rangle=0, \\
&& \frac{d\langle\chi_5\rangle}{dt}-\langle\chi_6\rangle
+\omega_{\perp}^2\langle\chi_4\rangle-i\frac{g}{\hbar}\langle n_0\chi_5\rangle
=\langle\chi_5\rangle_{\rm coll}, \\
&&\frac{d\langle\chi_6\rangle}{dt}
+2\omega_{\perp}^2\langle\chi_5\rangle+i\frac{g}{\hbar}
\left[\langle S_+\chi_6\rangle_0-\langle n_0\chi_6\rangle\right]
=\langle\chi_6\rangle_{\rm coll}, \nonumber \\ 
\end{eqnarray}
Just as in the previous case of the dipole mode, we truncate the hierarchy 
by assuming an appropriate form for the distribution
\begin{eqnarray}
\delta \sigma_+&=&f_0[\alpha_0 + \alpha_1 z^2 + \alpha_2 z p_z +
\alpha_3 p_z^2 \nonumber \\ 
&&+ \alpha_4 r_\perp^2 + \alpha_5 {\bf
r}_{\perp}\cdot{\bf p}_{\perp} + \alpha_6 p_\perp^2] .
\label{quadtruncate}
\end{eqnarray} 
The coefficients in the expansion can be related back to the set of
moments using Eq.~(\ref{momentdef}): $\alpha_0 = -m[\omega_z^2 \langle
\chi_1 \rangle + \omega_\perp^2 \langle \chi_4 \rangle + \langle
\chi_3 \rangle + \langle \chi_6 \rangle]/2 k_B T$,
$\alpha_1=(m\omega_z^2/k_BT)^2\langle \chi_1 \rangle/2$, $\alpha_2 = m
\omega_z^2 \langle \chi_2 \rangle /(k_B T)^2$, $\alpha_3 = \langle
\chi_3 \rangle/2 (k_B T)^2$, $\alpha_4=(m\omega_\perp^2/k_BT)^2\langle
\chi_4 \rangle/2$, $\alpha_5 = m \omega_\perp^2 \langle \chi_5 \rangle
/(k_B T)^2$, and $\alpha_6 = \langle \chi_6 \rangle/2 (k_B T)^2$.

Following the same procedure as for the dipole mode, and obtain a
closed set of coupled moment equations, written in matrix form as
\begin{equation}
{d \over {dt}} \chi = \hat W_q \chi,
\label{dtWq}
\end{equation}
where the coupling matrix $\hat W_q$ in this case is 
\begin{equation}
\label{Wq}
\hat{W}_{q}=-\left( \begin{array}{cccccc} 0 & -2 & 0 & 0 & 0 & 0\\
\omega _{z}^{2} & \tilde{\gamma }_{D}/2 & -1 & 0 & 0 & 0\\ 0 & 2\omega
_{z}^{2} & \tilde{\gamma }_{T}^z & 0 & 0 & \delta \gamma _{T}\\ 0 & 0 &
0 & 0 & -2 & 0\\ 0 & 0 & 0 & \omega _{\perp }^{2} & \tilde{\gamma
}_{D}/2 & -1\\ 0 & 0 & 2\delta \gamma _{T} & 0 & 2\omega _{\perp }^{2}
& \tilde{\gamma }_{T}^{\perp}
\end{array}\right).
\end{equation}
The vector of moments is ${\bf{\chi}}\equiv \{ \langle \chi_1 \rangle,
\langle \chi_2 \rangle, \cdots, \langle \chi_6 \rangle\}$. The tilde
on the relaxation rates indicates the complex form $\tilde \gamma =
(\gamma - i \omega_{\rm{MF}})$. The quantities $\gamma_T^z$ and
$\gamma_T^{\perp}$, given explicitly in the Appendix, are the spatially
averaged axial and radial thermal relaxation rates, the difference of
which $\delta \gamma = \gamma_T^{\perp} - \gamma_T^z$ is not zero in
general. We note that $\hat W_q$ is nearly block diagonal and that the
axial and radial oscillations are coupled only through collisions
associated with $\delta\gamma_T$.

By substituting $\chi=\chi_0 e^{-i\omega t}$ into Eq.~(\ref{dtWq}), we
obtain an eigenvalue equation with six solutions; the dispersion law
is determined from
\begin{equation}
\mathcal F_1^z(\omega)\mathcal F_1^\perp(\omega)
+\mathcal F_2(\omega)\mathcal F_3(\omega)=0,
\end{equation}
where
\begin{eqnarray}
&& \label{F1def}\mathcal F_1^i(\omega)=
\left(\omega^2-4\omega_i^2+\frac{2\omega_i^2}{1-i\omega/\tilde\gamma_{T}^i}
+i\omega\frac{\tilde\gamma_D}{2}\right) \\ &&\mathcal
F_2(\omega)=(\omega+i\tilde\gamma_{T}^z)(\omega+i\tilde\gamma_{T}^{\perp})
\\ &&\mathcal F_3(\omega)=2\delta\gamma_T^2
\left(\omega^2-2\omega_z^2+i\omega\frac{\tilde\gamma_D}{2}\right)
\nonumber \\
&&~~~~~~~~~~~~~~~~
\times\left(\omega^2-2\omega_{\perp}^2+i\omega\frac{\tilde\gamma_D}{2}\right),
\end{eqnarray}
In both the weak interaction
($\omega_i\gg|\tilde\gamma_D|,|\tilde\gamma_{T}^i|$) and strong
interaction ($\omega_i\ll|\tilde\gamma_D|,|\tilde\gamma_{T}^i|$)
limits, the axial and radial modes are uncoupled and are determined
from $\mathcal F_1^i(\omega)=0$, with $\mathcal F_1^i(\omega)$ given
in (\ref{F1def}). In the weak interaction limit, one has three modes
\begin{eqnarray}
\label{quadweak}\omega&=&2\omega_i-\frac{\omega_{\rm MF}}{2}-
\frac{i}{4}(\gamma_D+\gamma_T^i), \\
\omega&=&-\left(2\omega_i+\frac{\omega_{\rm MF}}{2}\right)
-\frac{i}{4}(\gamma_D+\gamma_T^i), \\
\omega&=&-\frac{\omega_{\rm MF}}{2}-\frac{i}{2}\gamma_T^i.
\end{eqnarray}
In the strong interaction limit, there is one (weakly damped)
low-frequency mode and two (strongly damped) high-frequency modes.
The low-frequency mode is given by
$\omega=-4i\omega_i^2/\tilde\gamma_D$, which can be written
approximately as
\begin{equation}
 \omega\simeq{4\omega_i^2 \over {\omega_{\rm{MF}}}} \Big ( 1 - 
{\gamma_D^2 \over \omega_{\rm{MF}}^2}-i
{\gamma_D \over \omega_{\rm{MF}}} \Big ) .
\label{quadstrong}
\end{equation}
%In the weak interaction limit, this goes over to (\ref{quadweak}).
This has the same scaling behavior with respect to the density,
scattering length and temperature as the dipole mode result in
Eq.~(\ref{dipolestrong}).

\section{Numerical solution of 1D spin kinetic equation}
In this section we compare the predictions of the moment method to a
direct numerical solution of the spin kinetic equation.  In
Refs.~\cite{Fuchs,Williams}, a one-dimensional model of the kinetic
equation was presented and found to give very good agreement with
experiments \cite{Lewandowski}.  A justification that was given for
the model is the separation of time scales between the axial and
radial directions. In hindsight, the success of the one dimensional
model can be further understood based on the moment method results in
the previous section that the axial and radial modes are uncoupled in
the linearized regime.

\begin{figure}
  \centerline{\epsfig{file=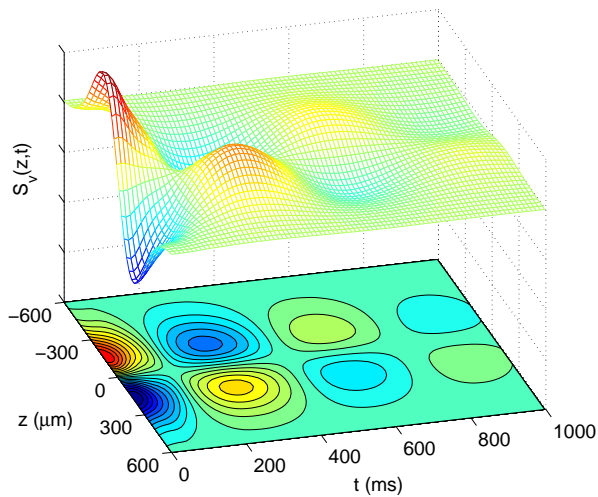,height=2.6in}}
\caption{Dipole excitation. The surface shows how the spin component
$S_v(z,t)$ varies with position and time. For visual clarity, we
also show the contours projected below the surface.}
\end{figure}
We construct a one-dimensional model of the system by making the
ansatz \(
\vec\sigma(\mathbf{r},\mathbf{p},t)=\vec\sigma(z,p,t)h_{0}({\bf r}_{\perp},
{\bf p}_{\perp})\) and then averaging over ${\bf r}_{\perp}$ and ${\bf p}_{\perp}$.
Here we take the static
profile in the radial direction to be of Gaussian form \( h_{0}=\exp
[-(p_{\perp}^{2}/2m+m\omega_{r}^{2}r_{\perp}^2/2)/k_{\mathrm{B}}T] \).
We substitute this
ansatz into (\ref{kineticfs}) and integrate over the radial phase
space variables, which gives the following one-dimensional model
Boltzmann equation
\begin{equation}
\label{kineticfs1D}
\frac{\partial \vec{\sigma }}{\partial t}+\frac{p}{m}\frac{\partial
\vec{\sigma }}{\partial z}-\frac{\partial U_{\mathrm{ext}}}{\partial
z}\frac{\partial \vec{\sigma }}{\partial p}-\vec{\Omega }_{n}\times
\vec{\sigma }=\left. \frac{\partial \vec{\sigma }}{\partial t}\right|
_{1D}.
\end{equation}
Here we have dropped terms that scale like $g n / k_B T$. The collision
integral in one dimension involves a phase space average in the radial
direction \( \left. \partial \vec{\sigma }/\partial t\right|
_{1D}\equiv \int _{xy}\left. \partial \vec{\sigma }/\partial t\right|
_{\mathrm{coll}}/\int _{xy}h_{0}, \) where we have introduced the
notation \( \int _{xy}\cdots \equiv \int d{\bf r}_{\perp}\int d{\bf p}_{\perp}
\cdots /(2\pi \hbar )^{2} \).
The radial averaging introduces a scaling factor in the mean field terms,
so that \( g\rightarrow g'=
g/(2\lambda_{\rm th}^2) \), where $\lambda_{\rm th}$ is the thermal de
Broglie wavelength. $g'$ has the correct units of energy times
distance required in our one dimensional model.

Although the direct numerical simulation using the full expression for
the one dimensional collision integral derived from
Eq.~(\ref{collision}) is technically feasible, we introduce a simple
model for the relaxation
\begin{equation}
\label{rel_approx}
\left.\frac{\partial \vec{\sigma}}{\partial t}\right|_{1D}
=-\frac{1}{\tau_{\rm cl}(z)}[\vec{\sigma}(z,p,t)
-\vec{M}(z,t)f_0(z,p)],
\end{equation}
where $\tau_\mathrm{cl}(z)=[16 a_{12}^2 n_0(z) \sqrt{\pi
k_{\mathrm{B}} T/m}]^{-1}$ is the radially averaged mean collision
time, $f_0(z,p)\equiv f_0(\mathbf{r},\mathbf{p})/h_0$, and
$\vec{M}(z,t)=\vec{S}(z,t)/n_0(z)$.  Equation (\ref{rel_approx}) contains
the essential properties of collisions: (i) it vanishes when the
distribution function has the local equilibrium form
$\vec{\sigma}(z,p,t) \propto \vec{M}(z,t)e^{-p^2/2mk_{\rm B}T}$, (ii)
it conserves the spin density. 
We note that the form Eq.~(\ref{rel_approx}) does not require the knowledge of
the long-time equilibrium solution for $\vec{S}({\bf r},t)$.
\begin{figure}
  \centerline{\epsfig{file=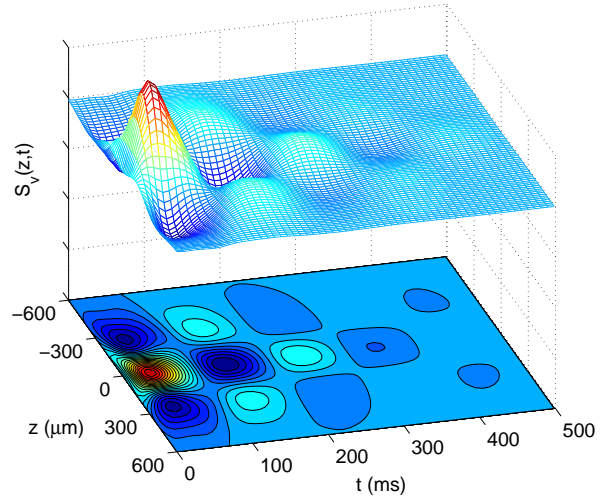,height=2.6in}}
\caption{Quadrupole excitation, as in Fig.~2}
\end{figure}

We solve Eq.~(\ref{kineticfs1D}) numerically using a finite difference,
alternating-direction Crank-Nicholson routine. As a check on the
numerics we monitor the integrated spin components $\int dz S_i(z,t)$,
which are conserved if $\vec \Omega = 0$, and the integrated spin
magnitude, which must satisfy the relation
\begin{equation}
{\partial \over {\partial t}} \int dz (S_x^2 + S_y^2 + S_z^2)
= 2 \int dz~ \vec J \cdot {d \vec S \over {dz}}.
\end{equation}
The total spin can decay to zero if there is an inhomogeneous field
$\vec \Omega(z,t)$ present (i.e. $\boldsymbol \nabla \vec \Omega \neq
0$).

To compare directly with the moment method results derived in the previous
section, we take as an initial state $\vec \sigma(z,t=0) = \vec
\sigma_0(z) + \delta \vec \sigma(z)$, where $\vec \sigma_0(z) = \{0,0,
f_0(z) \}$ and $\delta \vec \sigma(z) = \{ {\rm{Re}}\delta\sigma_+(z),
{\rm{Im}}\delta\sigma_+(z), 0 \}$. We take $\delta \sigma_+$ of the
form given in Eq.~(\ref{dipoletruncate}) for the dipole mode and
Eq.~(\ref{quadtruncate}) for the quadrupole mode.  The coefficients
$\alpha_i$ are determined by diagonalizing the coupling matrices
Eq.~(\ref{Wd}) and Eq.~(\ref{Wq}) respectively (in this section we consider
only the low frequency excitations for a given symmetry).

\begin{figure}
  \centerline{\epsfig{file=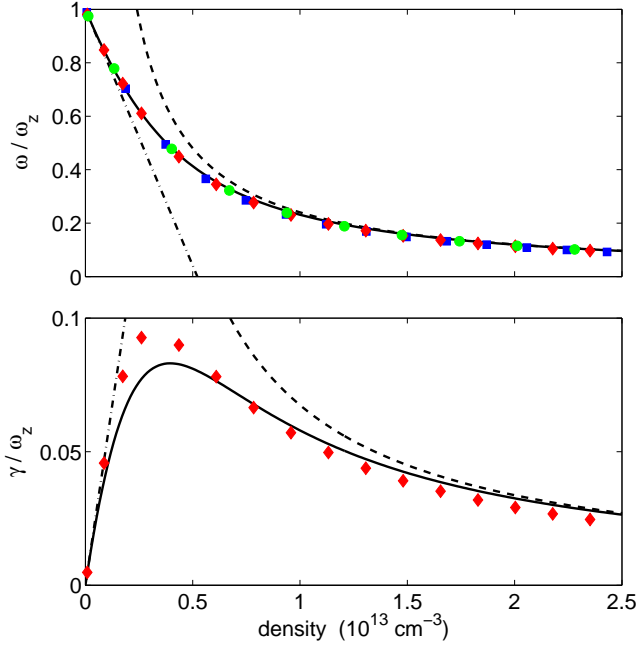,height=3.4in}}
\caption{Frequency and damping rate of the dipole mode versus peak
total density $n_0(0)$. The solid line is the low frequency solution
obtained from Eq.~(\ref{dipoledispersion}) and the chain and dashed
lines are the weak Eq.~(\ref{dipoleweak}) and strong
Eq.~(\ref{dipolestrong}) interaction limits, respectively. The filled
points are obtained from a direct numerical solution of
Eq.~(\ref{kineticfs1D}) for three different temperatures $T$: $600$ nK
(green circle), $800$ nK (red diamond), and $1$ mK (blue square).}
\end{figure}
In order to visualize the spatial form of the spin wave, we show the
transverse spin component $S_v(z,t)$ in Figures 2 and 3 as a function
of position and time to show the symmetry of the dipole and quadrupole
modes. We note the qualitative agreement between the contours of
Figure 3 and the graph shown in Figure 1b of Ref.~\cite{McGuirk}. We
do not plot the $S_u(z,t)$ component, which has the same structure but
is shifted $90^\circ$ out of phase in time. The transverse spin
$\{S_u(z,t), S_v(z,t) \}$ at a given position $z$ traces out a spiral
that terminates at the origin as $t\rightarrow \infty$, whose overall
diameter varies with position. The spin precession as a function of
position for the actual experiment is shown in Figure 2 of
Ref.~\cite{McGuirk}.

We extract a frequency and damping rate from the numerical solution by
calculating the dipole $\langle z \rangle$ or quadrupole $\langle z^2
\rangle$ moment and fitting this quantity to a damped sine function of
the form $A\exp(-B t)\sin(C t + D)$. The coefficients are obtained
using a least squares fitting routine. The frequency $\omega=C$ and
damping rate $\gamma=B$ are then compared to the predictions of the
moment method.
\begin{figure}
  \centerline{\epsfig{file=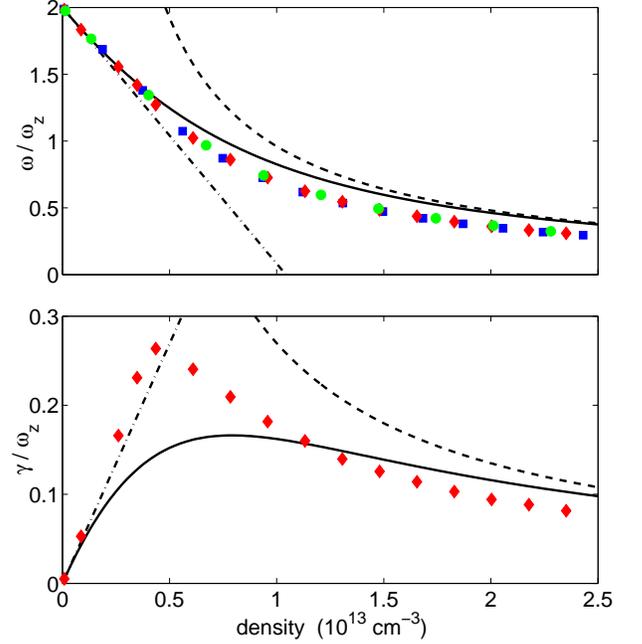,height=3.4in}}
\caption{Frequency and damping rate of the quadrupole mode versus peak
total density $n_0(0)$. The solid line is the low frequency solution
obtained from Eq.~(\ref{F1def}) (eg. $\mathcal F_1^z=0$) and the chain
and dashed lines are the weak Eq.~(\ref{quadweak}) and strong
Eq.~(\ref{quadstrong}) interaction limits, respectively. The filled
points are obtained from a direct numerical solution of
Eq.~(\ref{kineticfs1D}) for three different temperatures $T$: $600$ nK
(green circle), $800$ nK (red diamond), and $1$ mK (blue square).}
\end{figure}

In Figures 4 and 5 we plot the frequencies and damping rates for the
dipole and quadrupole spin waves. We take values for the physical
quantities corresponding to the JILA experiment, where $\omega_z/2\pi
= 7$ Hz, and $\omega_\perp/2\pi = 230$ Hz. We find that the
frequencies obtained from the numerical solution agree extremely well
with the moment method prediction for the dipole mode, while the
quadrupole mode shows only qualitative agreement.  We also find that
the frequencies in Fig.~4 and Fig.~5 are given as
temperature-independent functions of the peak density, which is
consistent with the moment results.  In general, one can show that the
streaming term of the linearized spin kinetic equation in
Eq.~(\ref{eq_sigmapm}) is invariant under the scale transformation
$T\to T',{\bf r}\to (T'/T)^{1/2}{\bf r}, {\bf p}\to(T'/T)^{1/2}{\bf
p}$ for given $n_0(0)$, and thus the frequency of any given mode is
temperature independent. In Figures 4 and 5 we do not show the damping
as a function of temperature, which scales approximately as $\sqrt{T}$
at fixed peak density, according to Eq.~(\ref{gammacl}).

The most interesting feature occurs in the damping, where we see that
the damping of the quadrupole mode is qualitatively different from
that predicted by the moment method. We postulate that this
difference, which is largest at intermediate densities, is due to
Landau-type damping arising from the mean-field coupling of the
collective mode to higher excitations.  Our moment approach using the
simple truncated form for the distribution function $\sigma_+$, as
given by Eq.~(\ref{dipoletruncate}) or Eq.~(\ref{quadtruncate}), does
not account for this effect \cite{Nikuni}.

For a homogeneous Bose gas, Oktel and Levitov worked out the spectrum
of spin waves in the collisionless regime \cite{Oktel2} using a
linear-response theory with the random phase approximation.  They find
that the mean-field coupling gives rise to Landau-type damping, which
is in addition to the collisional damping due to spin diffusion.  The
damping rate is given by
\begin{equation}
\gamma_L = {\sqrt{\pi} g^2 n^2 \over {\hbar^2 k v_{\rm{th}}}} 
e^{-(g n / \hbar k v_{\rm{th}})^2}.
\end{equation}
A rough estimate of Landau damping in a trapped gas may be given by
using $k\sim 1/R_{\rm th}$ in this uniform gas result.  However, this
simple estimate predicts a damping rate that is about an order of
magnitude larger than that shown in Figure 5. Apparently one needs to
work out the linear response calculation taking into account the
excitation spectrum explicitly in a trap potential in order to obtain
a better quantitative model for Landau damping.

In Ref.~\cite{McGuirk}, we compare the numerical solution presented in
this paper to experimental data, and find excellent agreement for both
frequency and damping. It is important to realize that in our
numerical calculation, we have used a relaxation time approximation to
treat the collision integral. In this model, one has some freedom to
choose an appropriate relaxation time; there are a few different
candidates, but here we have used the spatially dependent form of the
mean collision time given in Eq.~(\ref{rel_approx}). Although this is
a reasonable choice, it was not {\emph{a priori}} obvious that it
would result in the best agreement with experimental data.

We also remark that our results are strictly valid for the case where
the inhomogeneity $\Delta(\mathbf{r})$ is vanishingly small (though we
have not carried out the full linear response theory, we envision
$\Delta(\mathbf{r})$ as playing the role of an external perturbation
used to excite the spin wave). In the JILA experiment~\cite{McGuirk},
the effect of applying a constant static $\Delta(\mathbf{r})$ during
the entire spin wave oscillation was investigated as a function of the
magnitude of the perturbation. For a large inhomogeneity, the response
of the system is nonlinear and the mode frequency is modified; the
earlier JILA experiment~\cite{Lewandowski} seems to reside in this
regime.

\section{Conclusion}

In this paper we have studied spin waves in a dilute noncondensed Bose
gas of two level atoms. Our main new contribution is that we have
treated an inhomogeneous system held in a harmonic trap in order to
describe related experiments on spin waves~\cite{McGuirk}. We applied
the moment method technique for dilute trapped gases, which lead to
closed form solutions for the frequencies and damping of dipole and
quadrupole modes. As a test of the validity of our moment model, we
compared the results to the numerical calculation of a one dimensional
model of the spin Boltzmann equation and found very good agreement
overall. The main discrepancy between the moment model and the
numerical calculation occurred in the damping of the quadrupole mode.
We attribute this to the Landau damping, which is contained in the
numerical approach, but absent in the moment model. 

Although the salient features of spin waves in a noncondensed dilute
gas seem to be well described by our theory, which shows excellent
agreement with experimental data presented in Ref.~\cite{McGuirk}, the
situation for this system just below the Bose-Einstein condensation
temperature~\cite{NW} remains largely unexplored.  Many {\emph{zero
temperature}} properties of a spin-1/2 condensate have been studied,
when the thermal component is absent, such as spin
winding~\cite{Matthews99} and vortex-like spin
textures~\cite{Williams99,Matthews99b}. It should be noted, however,
that at zero temperature the condensate does not support spin waves of
the type considered in this paper, since the exchange mechanism does
not occur in the condensate. It is very interesting that the spin-1/2
thermal gas exhibits strong collective behavior in the spin dynamics
due to the exchange effect, even in the collisionless regime; this is
in sharp contrast to the behavior of density fluctuations of a single
component thermal gas, where the mean field of the noncondensate plays
a very minor role (mainly as a source of damping of the condensate
excitations)~\cite{ZNG,Jackson}. At finite temperatures, the spins of
the condensate and thermal gas will interact strongly. For future
studies, it will be interesting to investigate how long-lived spin
textures in the condensate are affected by the thermal gas, which in
principle can itself support (probably short-lived) spin textures. It
will also be important to understand how the condensate modifies spin
waves in the thermal cloud~\cite{Oktel3}.

\section*{Acknowledgements}
We would like to thank E. A. Cornell for inspiring us to study the
linearized spin waves and to the JILA team of H. J. Lewandowski,
J. M. McGuirk, and D. M. Harber for all of their insight as well as
providing us with experimental results. We also thank D. Feder and
J. Roberts for proofreading the manuscript.

\begin{appendix}

\section*{APPENDIX: Collision integrals and relaxation times}

The collision integrals appearing in the kinetic equations
Eq.~(\ref{kineticfcm}) and Eq.~(\ref{kineticfs}) for the center of
mass and spin distribution functions are given by
\begin{eqnarray}
\left.\frac{\partial f}{\partial t}\right|_{\rm coll}
&=&\frac{\pi g^2}{\hbar}
\int \frac{d{\bf p}_2}{(2\pi\hbar)^3}\int \frac{d{\bf p}_3}{(2\pi\hbar)^3}
\int d{\bf p}_4 \nonumber \\
&&\delta(\epsilon_p+\epsilon_{p_2}+\epsilon_{p_3}+\epsilon_{p_4})
\delta({\bf p}+{\bf p}_2-{\bf p}_3-{\bf p}_4) \cr
&&\times \{ 3[f({\bf p}_3) f({\bf p}_4)-f({\bf p}) f({\bf p}_2)]
\nonumber \\ &&+\vec{\sigma}({\bf
p}_3)\cdot\vec{\sigma}({\bf p}_4) -\vec{\sigma}({\bf
p})\cdot\vec{\sigma}({\bf p}_2) \}.
\end{eqnarray}
\begin{eqnarray}
\left.\frac{\partial\vec{\sigma}}{\partial t}\right|_{\rm coll}
&=&\frac{\pi g^2}{\hbar} \int \frac{d{\bf p}_2}{(2\pi\hbar)^3}\int
\frac{d{\bf p}_3}{(2\pi\hbar)^3} \int d{\bf p}_4 \cr
&&\delta(\epsilon_p+\epsilon_{p_2}-\epsilon_{p_3}-\epsilon_{p_4})
\delta({\bf p}+{\bf p}_2-{\bf p}_3-{\bf p}_4) \cr
&&\times\{3f({\bf p}_3)\vec{\sigma}({\bf p}_4)
+\vec\sigma({\bf p}_3)f({\bf p}_4) \cr
&&-f({\bf p})\vec{\sigma}({\bf p}_2)
-3\vec{\sigma}({\bf p})f({\bf p}_2)\},
\label{collision}
\end{eqnarray}
where $\epsilon_p\equiv p^2/2m$. Here we neglect a principal value
contribution, which gives a second-order
correction to the free streaming evolution, and
we take all scattering lengths $a_{ij}$ to be equal
- a reasonable approximation for ${}^{87}$Rb. This approximation 
results in the conservation of spin density
during collisions, i.e. $\int d{\bf p} \partial\vec{\sigma}/
\partial t|_{\rm coll}=0$. When the small differences in scattering
lengths are accounted for, the transverse spin decays 
slowly. For ${}^{87}$Rb, this
contribution to the ``T2'' lifetime is of the order of
10 s \cite{NW}.

For the linearized form of the spin collision term appearing in
Eq. (\ref{linear_eq}), it is convenient to express the fluctuation
of the spin distribution function as
\begin{equation}
\delta\vec{\sigma}({\bf r},{\bf p},t)=f_0({\bf r},{\bf p})
\vec{\psi}({\bf r},{\bf p},t).
\end{equation}
Then the linearized collision integral is written as
$\left.\partial\vec{\sigma}/\partial t\right|_{\rm coll}\equiv
L[\vec{\psi}]$
\begin{eqnarray}
 L[\vec{\psi}]
&=&\frac{\pi g^2}{\hbar}
\int \frac{d{\bf p}_2}{(2\pi\hbar)^3}\int \frac{d{\bf p}_3}{(2\pi\hbar)^3}
\int d{\bf p}_4 \nonumber \\ 
&\times& \delta(\epsilon_p+\epsilon_{p_2}-\epsilon_{p_3}-\epsilon_{p_4})
\delta({\bf p}+{\bf p}_2-{\bf p}_3-{\bf p}_4) \nonumber \\
&\times& f_0({\bf p}_3)f_0({\bf p}_4) \{
2[\vec{\psi}({\bf p}_4)-\vec{\psi}({\bf p})] \nonumber \\
&+&[\vec{\psi}({\bf p}_4)+\vec{\psi}({\bf p}_3)-\vec{\psi}({\bf p}_2)
+\vec{\psi}({\bf p})]\}.
\end{eqnarray}

Using the approximate forms of the distribution function Eq.~(\ref{dipoletruncate}) and
Eq.~(\ref{quadtruncate}) in the linear collision operator and taking moments, 
we find that the collisional contributions to the moment equations 
are given by
\begin{equation}
\langle p_{x_i} \rangle_{\rm coll}=-\gamma_D\langle p_{x_i}\rangle,
\end{equation}
\begin{equation}
\langle zp_z \rangle_{\rm coll}=-\frac{\gamma_D}{2}\langle zp_z\rangle,
\end{equation}
\begin{equation}
\langle {\bf r}_{\perp}\cdot{\bf p}_{\perp} \rangle_{\rm coll}
=-\frac{\gamma_D}{2}\langle {\bf r}_{\perp}\cdot{\bf p}_{\perp}\rangle,
\end{equation}
\begin{equation}
\langle p_z^2 \rangle_{\rm coll}=-\gamma_T^z \langle p_z^2\rangle
-\delta\gamma_T \langle p_{\perp}^2\rangle,
\end{equation}
\begin{equation}
\langle p_{\perp}^2 \rangle_{\rm coll}=-\gamma'_T \langle p_{\perp}^2\rangle
-2\delta\gamma_T \langle p_z^2\rangle.
\end{equation}
The various relaxation rates are given by spatial average of the
following spin transport relaxation times:
\begin{equation}
\frac{n_0({\bf r})}{\tau_D({\bf r})}
\equiv-\frac{1}{3mk_{\rm B}T}\int\frac{d{\bf p}}{(2\pi\hbar)^3}
{\bf p}L[{\bf p}],
\end{equation}
\begin{equation}
\frac{n_0({\bf r})} {\tau_T({\bf r})}\equiv
-\int \frac{d{\bf p}}{(2\pi\hbar)^3}\frac{1}{2(mk_{\rm B}T)^2}p_z^2L[p_z^2],
\end{equation}
\begin{equation}
\frac{n_0({\bf r})}{\tau_T^{\perp}({\bf r})}\equiv
-\int \frac{d{\bf p}}{(2\pi\hbar)^3}\frac{1}{4(mk_{\rm B}T)^2}
p_{\perp}^2L[p_{\perp}^2].
\end{equation}
The associated spatially averaged relaxation rates are 
given by
\begin{equation}
\gamma_D\equiv \frac{1}{N}\int d{\bf r}\frac{n_0({\bf r})}{\tau_D({\bf r})}
=\frac{1}{2\sqrt{2}}\frac{1}{\tau_D(0)}.
\end{equation}
\begin{equation}
\gamma_T^z\equiv\frac{1}{N}\int d{\bf r}\frac{n_0({\bf r})}{\tau_T({\bf r})}
=\frac{1}{2\sqrt{2}\tau_T^z(0)},
\end{equation}
\begin{equation}
\gamma_T^{\perp}\equiv\frac{1}{N}\int d{\bf r}\frac{n_0({\bf r})}{\tau'_T({\bf r})}
=\frac{1}{2\sqrt{2}\tau_T^{\perp}(0)},
\end{equation}

For detailed calculations of various transport relaxation times
in a trapped Bose-condensed gas, we refer to Ref.~\cite{LK}.
It is straightforward to generalize those calculations to work out
the spin transport relaxation times defined above.
We find
\begin{eqnarray}
\tau_D^{-1}({\bf r})&=&\frac{1}{3}\tau_{\rm cl}^{-1}({\bf r}), \\
{\tau_T^z}^{-1}({\bf r})&=&\frac{3}{5}\tau_{\rm cl}^{-1}({\bf r}), \\
{\tau_T^{\perp}}^{-1}({\bf r})&=&\frac{7}{15}\tau_{\rm cl}^{-1}({\bf r}),
\end{eqnarray}
where $\tau_{\rm cl}({\bf r})\equiv [32n_0({\bf r})a^2(\pi k_{\rm B}T/m)^{1/2}]^{-1}$
is the mean-collision time.
The spatially averaged relaxation rates are given by
\begin{equation}
\gamma_D=\frac{1}{3}\gamma_{\rm cl},~~
\gamma_T^z=\frac{3}{5}\gamma_{\rm cl},~~
\gamma_T^{\perp}=\frac{7}{15}\gamma_{\rm cl},~~
\delta\gamma_T=-\frac{2}{15}\gamma_{\rm cl},
\end{equation}
where $\gamma_{\rm cl}$ is the spatially averaged mean collision rate
\begin{equation}
\gamma_{\rm cl}=\frac{16}{\sqrt{2}}n_0(0)a^2\left(\frac{\pi k_{\rm B}T}{m}
\right)^{1/2}.
\label{gammacl}
\end{equation}

In order to compare the moment method directly with our numerical solution of the 1D
kinetic equation, we also worked out the three relaxation rates within
the simple relaxation time approximation
\begin{equation}
\left.\frac{\partial\vec{\sigma}}{\partial t}\right|_{\rm coll}
=-\frac{\vec{\sigma}({\bf r},{\bf p},t)-\vec{M}({\bf r},t)
f_0({\bf r},{\bf p})}{\tau_{\rm cl}({\bf r})}.
\label{relax_approx}
\end{equation}
This simple formula leads to the same collisional contributions in the
moment equations as given from the original collision integral, with
all the transport relaxation times being replaced with $\tau_{\rm
cl}({\bf r})$, so that one has
$\gamma_D=\gamma_T=\gamma_T'=\gamma_{\rm cl}$.  Thus, within the
relaxation time approximation, oscillations in the axial and radial
directions are completely uncoupled since $\delta\gamma_T=0$.

\end{appendix}

\bibliography{spin}

\end{document}